\documentclass[prc,aps,twocolumn,preprintnumbers,amsmath,amssymb,showpacs,floatfix]{revtex4}
\usepackage{amsmath}
\usepackage{graphicx}
\usepackage{floatflt,epsfig}
\usepackage{lscape}
\def\eeq{\end{equation}}
\def\beqa{\begin{eqnarray}}
\def\eeqa{\end{eqnarray}}
\def\ban{\begin{eqnarray*}}
\def\ean{\end{eqnarray*}}
\def\bi{\begin{itemize}}
\def\ei{\end{itemize}}

\begin{document}
\title{Microscopic calculations of double and triple Giant Resonance
excitation in heavy ion collisions}
\author{E. G. Lanza}
\affiliation{I.N.F.N.-Catania and Dipartimento di Fisica e Astronomia,
Universit\'a di Catania,  Via S. Sofia 67, I-95123 Catania, Italy}
\author{M. V. Andr\'es}
\affiliation{Departamento de F\'{i}sica Atomica, Molecular y Nuclear,
Universidad de Sevilla, Apdo 1065, E-41080 Sevilla, Spain}
\author{F. Catara}
\affiliation{I.N.F.N.-Catania and Dipartimento di Fisica e Astronomia,
Universit\'a di Catania,  Via S. Sofia 67, I-95123 Catania, Italy}
\author{Ph. Chomaz}
\affiliation{GANIL (DSM-CEA/IN2P3-CNRS), B.P. 55027, F-14076 Caen C\'edex 5, France}
\author{M. Fallot}
\affiliation{Subatech, 4 rue Alfred Kastler BP 20722, F-44307 Nantes
Cedex 3, France}
\author{J. A. Scarpaci}
\affiliation{Institut de Physique Nucl\'{e}aire, IN2P3-CNRS, F-91406
Orsay Cedex, France}

\begin{abstract}
  We perform microscopic calculations of the inelastic cross sections
  for the double and triple excitation of giant resonances induced by
  heavy ion probes within a semicalssical coupled channels
  formalism. The channels are defined as eigenstates of a bosonic
  quartic Hamiltonian constructed in terms of collective RPA
  phonons. Therefore, they are superpositions of several multiphonon
  states, also with different numbers of phonons and the spectrum is
  anharmonic. The inclusion of (n+1) phonon configurations affects the
  states whose main component is a n-phonon one and leads to an
  appreacible lowering of their energies. We check the effects of such
  further anharmonicities on the previous published results for the
  cross section for the double excitation of Giant Resonances. We find
  that the only effect is a shift of the peaks towards lower energies,
  the double GR cross section being not modified by the explicity
  inclusion of the three-phonon channels in the dynamical
  calculations. The latters give an important contribution to the
  cross section in the triple GR energy region which however is still
  smaller than the experimental available data. The inclusion of four
  phonon configurations in the structure calculations does not modify
  the results.

\end{abstract}


\maketitle

\vspace{0.50cm}

\section{Introduction}
\smallskip 

Since their discovery (about 70 years ago) the Giant Resonances have
been considered as the best example of coherent motion of nuclear
systems\cite{book}. From a macroscopic point of view, this collective
behavior can be considered as high frequency, harmonic vibrations of
the nuclear density around its equilibrium shape. If these
oscillations were harmonic then higher states of equidistant energies
should exist. The first experimental indication of some structures in
the excitation function in heavy ion collisions which could be
interpreted as due to the population of Double Giant Resonances dates
back to 1977 (see reviews in ref.\cite{rev} and references
therein). Recently there has been an unambiguous evidence of the
existence of a Double Giant Quadrupole Resonance\cite{sca}. More
recently also the Triple Giant Quadrupole Resonance has been observed
at GANIL by using, for the first time, the SPEG spectrometer in
conjunction with the INDRA 4$\pi$ detector\cite{fra,sca1}. Multiphonon
excitations were clearly observed in double charge exchange reactions
using ($\pi^+,\pi^-$) and ($\pi^-,\pi^+$) reactions\cite{mor}. The
study of the excitation of the double GDR of very heavy nuclei by
means of relativistic Coulomb excitation have been investigated in
experiments performed at GSI within the LAND
collaboration\cite{land}. These are exclusive experiments where
projectile fragments, neutron and gamma rays from the excited
fragments are measured. More recently, the same group has found
a hint for a three-phonon dipole state by measuring differential
cross section for electromagnetic fission of $^{238}$U at relativistic
energy\cite{land-3}.

The theoretical studies to have a better comprehension of the
multiphonon problematic have taken various directions. An extension of
the quasi-particle Random Phase Approximation (RPA) has been used in
ref. \cite{pon} where, in addition to the mixing of one- and
two-phonon states, also some specific three-phonon configurations have
been considered as a mechanism to generate the damping width of the
DGDR. Other approaches exploit the so called Brink-Axel
hypothesis\cite{bri}: A Giant Resonance can be excited on top of any
nuclear excited state. In this approach\cite{carl,wei}, the states to
which the one-phonon states decay are described in terms of GOE
(Gaussian Orthogonal Ensemble). By means of the Random Matrix Theory
the average cross section for the excitation of the double GDR is
calculated as a function of the spreading and damping width. The
inelastic cross sections of the double and triple giant dipole
resonances in a Coulomb excitation process\cite{hus} have been also
calculated.

The excitation of collective vibrational states in heavy ion
collisions can be viewed as due to the action of the mean field of
each collision partner on the other. A rather good microscopic
description of such states is given by Random Phase Approximation. It
can be introduced as the lowest order in a boson expansion leading to
a boson image of the Hamiltonian which is a sum of independent
harmonic oscillators corresponding to each collective mode. Thus the
RPA states are pure one-phonon and multiphonon states and their energy
is the sum of the energies of the single phonons. When further terms
in the boson expansion are taken into account, anharmonicities arise
and the eigenstates of the Hamiltonian are superpositions of
multiphonon states, also with different numbers of phonons.

The mean field $U$ of each nucleus is a one body operator. When the
ground state of each nucleus is approximated by the uncorrelated
Hartree-Fock (HF) one, only the particle-hole (ph) part of $U$
enters. Its bosonic image is linear in the phonon creation and
annihilation operators. Non linear terms arise in a natural way when
the correlations in the ground state are taken into account and thus
the particle-particle and hole-hole parts of $U$ are no more
negligible. Indeed, the bosonic images of such terms are non-linear
in the RPA phonons.

In several previous papers\cite{vol1,lan1,and}, it has been shown that
the above mentioned anharmonicities and non-linearities bring to a
much  better agreement between theoretical and experimental cross
sections for the excitation of double giant resonance states in heavy
ion collisions. 

Till now only the cross sections to one- and two-phonon states have
been calculated within the above recalled approach. (For simplicity,
we will call ``n-phonon state'' an eigenstate of the Hamiltonian whose
main component is a n-phonon configuration). In ref.\cite{fal},
however, we have computed the vibrational spectrum of $^{40}$Ca and
$^{208}$Pb by extending the previous calculations with the inclusion
of three-phonon states. It was found that the two-phonon states are
affected by this extension. In particular, the double Giant Resonances
(GR) are appreciably pushed down. The lesson we learn from this is
that one cannot perform calculations on the double GR without taking
into account the effect of the three-phonon states. In view of that,
in the present paper we recalculate the cross section for the
inelastic scattering $^{40}$Ca + $^{40}$Ca at 50 MeV/A and $^{208}$Pb +
$^{208}$Pb at 641 MeV/A in the enlarged space in order to check how
much our previous results on the cross section in the double GR region
are modified. Then we will present calculations of the inelastic cross
sections for the triple excitation of the GR where we include the
three-phonon states. In these latter calculations we will also take into
account the effects of the four-phonon states.


\section{Approaches and formalism}
\smallskip 

In this section we shortly recall our approach and how anharmonicities
in the spectrum and non-linearities in the excitation operator are
treated in it. Our model makes use of standard semiclassical methods
techniques. These methods are based on the assumption that nuclei move
on classical trajectories, while the internal degrees of freedom are
treated quantum mechanically.


\subsection{Multiphonon structure and anharmonicities}
\smallskip 

Let us denote by p (h) the single particle states which are
unoccupied (occupied) in the HF ground state of the nucleus and
introduce the mappings\cite{lamb}
\begin{equation}\label{map1}
 a^{\dagger}_p a_{h} \rightarrow B_{ph}^{\dagger}+(1- \sqrt{2}) ^{}
\sum_{p'h'}B^{\dagger}_{p'h'}B^{\dagger}_{p'h}B_{ph'}+~... \> ,
\end{equation}
and 
\begin{equation}\label{map2}
 a^{\dagger}_p a_{p'} \rightarrow
\sum_{h}B^{\dagger}_{ph}B^{\dagger}_{p'h} \> ; \> \> \>
 a^{\dagger}_h a_{h'} \rightarrow
\sum_{p}B^{\dagger}_{ph}B^{\dagger}_{ph'} \>,
\end{equation}
where $B^{\dagger}_{ph}$ and $B_{ph}$ are bosonic operators
\begin{equation}
[B_{ph},B^{\dagger}_{p'h'}]=\delta_{pp'} \delta_{hh'} \>.
\end{equation}
The second term in the right-hand side of eq. (\ref{map1}) is a
correction taking care of the Pauli principle. The fermionic
Hamiltonian is then mapped onto
\begin{eqnarray}
&&H_{B}= (H_{10}B^{\dagger }+H_{11}B^{\dagger }B+H_{20}B^{\dagger}
B^{\dagger})+ h.c. + \\  \nonumber
&&(H_{21}B^{\dagger}B^{\dagger }B+
H_{22}B^{\dagger }B^{\dagger }BB +
H_{31}B^{\dagger}B^{\dagger}B^{\dagger }B)+h.c.
\label{hb} 
\end{eqnarray}
where we have dropped indices for simplicity. The term $H_{10}$
vanishes in the HF basis. Collective phonon operators are introduced
by means of the Bogoliubov transformation
\begin{equation}
Q_{\nu }^{\dagger }= \sum_{p,h}(X_{ph}^{\nu }B_{ph}^{\dagger}-
Y_{ph}^{\nu }B_{ph}) \> .  
\label{q}
\end{equation}
The $X$ and $Y$ coefficients can be chosen so that the part of the
Hamiltonian which is quadratic in the $B^{\dagger}$ and $B$ operators
is diagonal when expressed in terms of the $Q^{\dagger}$ and $Q$ ones
\begin{equation}\label{rpa}
 H_{RPA}=\sum_{\nu }E_{\nu }Q_{\nu }^{\dagger}Q_{\nu }  
\end{equation}
and the $X$ and $Y$ satisfy the RPA equations. Of course, the spectrum
of $ H_{RPA}$ is harmonic. The other terms of the bosonic Hamiltonian
(\ref{hb}) introduce anharmonicities since they mix multiphonon states
among themselves. In our model we neglect the $H_{31}$ term because it
is smaller than the others, as it has also been checked in an extended
Lipkin model\cite{vol2}. For the remaining terms we keep only
\begin{eqnarray}
&&H_{21}B^{\dagger}B^{\dagger }B + h.c. \rightarrow
{\mathbb H_{21}}Q^{\dagger}Q^{\dagger }Q + h.c. \>\> ,\\
\nonumber {\rm and}&&\\
&&H_{22}B^{\dagger }B^{\dagger }BB \rightarrow
{\mathbb H_{22}}Q^{\dagger}Q^{\dagger }QQ \>\> ,
\end{eqnarray}
because the others are smaller by a factor $Y/X$ or powers of
it. Therefore our bosonic Hamiltonian becomes
\begin{eqnarray}
{\mathbb H}_{Q} &=&\sum_\nu E_\nu Q^{\dagger}_\nu Q_\nu +
\sum_{\nu_1 \nu_2 \nu} {\mathbb H}_{21} Q^{\dagger}_{\nu_1}
Q^{\dagger}_{\nu_2} Q_\nu + h.c. \nonumber\\
&+&\sum_{\nu_1 \nu_2 \nu_1^\prime
\nu_2^\prime} {\mathbb H}_{22} Q^{\dagger}_{\nu_1} Q^{\dagger}_{\nu_2}
Q_{\nu_1^\prime} Q_{\nu_2^\prime}  \> .
\label{h4} 
\end{eqnarray}
The eigenstates and eigenvalues of ${\mathbb H}_Q$ are then found by
diagonalizing it in the space of the states containing up to a certain
number $N_{\rm pho}$ of phonons. The ${\mathbb H}_{22}$ term mixes
multiphonon states with the same number of phonons, while the
${\mathbb H}_{21}$ mixes states having number of phonons differing by
one. In ref.\cite{fal} we have shown the results for $^{40}$Ca and
$^{208}$Pb obtained in the space with $N_{pho}$=3 and we have compared
them with those of ref.\cite{lan1} where we had $N_{pho}$=2. In the
next section we will mention the main results obtained there.


\subsection{Non linearities in the excitation operator}
\smallskip 

Within a semiclassical approach to nucleus-nucleus collisions the
excitation of one partner (say A) is due to the mean field of the
other (B) acting on it. Therefore the excitation operator is 
\begin{equation}
W (t)= \sum_{ij} W_{ij}(t) a^\dagger_i a_j
\end{equation}
where $W_{ij}(t)= (i|U_B(\vec R(t))|j)$ and $U_B$ is the mean field of
$B$ (including Coulomb) which depends on time through the relative
distance between the two nuclei. The indices $i$ and $j$ denote single
particle states, both occupied and unoccupied, in nucleus A. When one
neglects the correlation present in the ground state, only the $ph$
terms of $W$ are effective and the boson image of $W$ is linear in the
$Q^\dagger$ and $Q$ operators. Taking into account the correlations
the $pp$ and $hh$ parts of $W$ cannot be neglected and this brings to
quadratic terms
\begin{eqnarray}\label{WW} 
W = W^{00} &+& (\sum_\nu W^{10}_\nu Q^\dagger_\nu +
\sum_{\nu\nu^\prime} W^{11}_{\nu\nu^\prime} Q^\dagger_\nu
Q_{\nu^\prime} \\ \nonumber
&+& \sum_{\nu\nu^\prime} W^{20}_{\nu\nu^\prime}
Q^\dagger_\nu Q^\dagger_{\nu^\prime}) + h.c.  
\end{eqnarray}
The first term in this equation represents the interaction of the two
colliding nuclei in their ground state. The $W^{10}$ part connects
states differing by one phonon, the $W^{11}$ term couples excited
states with the same number of phonons, while $W^{20}$ allows
coupling between states differing by two phonons. These new
routes of excitation may increase the excitation probability of the
multiple GR.


\subsection{The cross section}
\smallskip 
The inelastic scattering cross section is calculated within a
semiclassical coupled channel approach. Let $|\Phi_\alpha>$ denote the
excited state of the nucleus of which we want to calculate the
excitation probability. These states are eigenstates of the Hamiltonian
(\ref{h4}) and therefore are superpositions of multiphonon states
obtained by diagonalizing ${\mathbb H}_Q$ in the space containing up
to $N_{pho}$ phonon configurations. The excitation probability
amplitudes satisfy, for each impact parameter $b$, the set of coupled
differential equations
\begin{equation} 
\dot A_\alpha (t) = -i \sum_{\alpha^\prime} e^{i (E_\alpha
    - E_{\alpha^\prime}) t} <\Phi_\alpha|W(t)|\Phi_{\alpha^\prime}> 
      A_{\alpha^\prime} (t)
\label{chan}
\end{equation}
which are integrated along classical trajectories with various impact
parameters $b$\cite{lan1,and}. The cross section to excite the state
$|\Phi_\alpha>$ is then calculated as 
\begin{equation}
\sigma_\alpha = 2\pi \int_{0}^{+\infty} P_\alpha (b) T(b) b db.
\label{xsec}
\end{equation}
where $ P_\alpha(b) = |A_\alpha (b,t = + \infty)|^2 $. The integral is
over the whole impact parameters range which is modulated by the
transmission coefficient $T(b)$.


\section{Excitation of $^{40}$Ca }

In this section we briefly recall the results we have obtained for the
structure calculation for the $^{40}$Ca\cite{fal}. As stated also in
the introduction, in that calculations we were guided by the results
we got from a study of an extended Lipkin-Meshow-Glick (LMG)
model\cite{vol2}.  In that paper, the original LMG model has been
extended in order to include terms that play the same role than the
anharmonic terms of our Hamiltonian (\ref{h4}). The Hamiltonian of
such extended LMG model is still exactly solvable.  The relevant
results can be summarized as follows: its diagonalization in an
enlarged space including up to three-phonon states produces results
which are very close to the exact ones\cite{vol2}. Therefore we have
followed this approach to calculate the spectrum of $^{40}$Ca in the
space of one-, two- and three-phonon states.

We have used a discrete selfconsistent HF+RPA with a SGII interaction,
including all one-phonon states with J$\leq$3 which exhaust at least
5\% of the EWSR. For the nucleus $^{40}$Ca, we have used the nine
one-phonon basis shown in table~\ref{basis-ca}. We have constructed
all two- and three-phonon configurations out of them, without energy
cut-off, with both natural and unnatural parity. Then the Hamiltonian
(\ref{h4}) has been diagonalized in the space spanned by such
states. The eigenstates are mixed states whose components are of one-,
two- and three-phonon kind
\begin{eqnarray}
\label{mix-3} \nonumber
|\Phi_\alpha>  &=& \sum_{\nu_1} c^\alpha_{\nu_1} |\nu_1> + 
\sum_{\nu_1\nu_2} c^\alpha_{\nu_1\nu_2} |\nu_1 \nu_2> \\ 
&+& \sum_{\nu_1\nu_2\nu_3} c^\alpha_{\nu_1\nu_2\nu_3} 
|\nu_1\nu_2\nu_3>  \> .
\end{eqnarray}

\begin {table} 
\caption {\label{basis-ca} RPA one-phonon basis for the nucleus
$^{40}$Ca. For each state, spin, parity, energy and percentage of the
EWSR(isovector for the GDR and the IVGQR and isoscalar for all the
other states) are reported. In the last two columns we report the
energies of the phonons after the inclusion of two and three-phonon
states, respectively.}\smallskip
\begin{center}
\begin{tabular}{|l|crrrr|}
\hline 
State&$J^\pi$&$E_{harm}$ & $EWSR$&$E_{2ph}$&$E_{3ph}$\\ 
& &$(MeV)$ &  $(\%)$&$(MeV)$&$(MeV) $\\ 
\hline  
GMR$_1$&$ 0^+$&$ 18.25$&$ 30$&$18.36$&$18.30$\\
GMR$_2$&$ 0^+$&$ 22.47$&$ 54$&$22.00$&$21.78$\\ \hline  
GDR$_1$&$ 1^-$&$ 17.78$&$ 56$&$17.35$&$17.29$\\
GDR$_2$&$ 1^-$&$ 22.03$&$ 10$&$21.64$&$21.59$\\ \hline  
ISGQR  &$ 2^+$&$ 16.91$&$ 85$&$16.51$&$16.44$\\
IVGQR  &$ 2^+$&$ 29.59$&$ 26$&$29.09$&$29.00$\\ \hline
$3^-$  &$ 3^-$&$ ~4.94$&$ 14$&$~4.47$&$~4.40$\\
LEOR   &$ 3^-$&$ ~9.71$&$ ~5$&$~9.33$&$~9.28$\\
HEOR   &$ 3^-$&$ 31.33$&$ 25$&$30.80$&$30.89$\\ \hline
\end{tabular}
\end{center}
\end{table}

The inclusion of the three-phonon states changes the energies of the
phonon basis of a few hundred of KeV as it is shown in
table~\ref{basis-ca}. The main result of the calculation is that the
spectrum of the two-phonon states is strongly modified by their
coupling to the three-phonon ones. The diagonalization in the
three-phonon space produces very large shifts in the energies, more
than one MeV (for $^{40}$Ca) in almost all the cases and always
downwards.  In view of these results, the inelastic scattering cross
section to two-phonon states has to be recalculated and compared with
that obtained without the inclusion of three-phonon
configurations\cite{and}.


\subsection{\protect\smallskip Cross section at 50 MeV/A}

The semiclassical calculations for the reaction $^{40}Ca$+$^{40}Ca$ at
50 MeV/A have been performed within the framework described above. In
this section we discuss the inelastic cross section calculations done
by including up to the three-phonon states.

In the case we are interested in, the nuclear contribution is
important. The form factors have been calculated\cite{lan2} by
employing a double folding procedure with the transition densities
calculated within the RPA. Furthermore, it has been introduced an
optical potential in order to avoid the uncertainty on the integration
over the impact parameters.  Since the optical potential takes into
account the absorption due to all channels, we have introduced a
procedure in order to avoid double counting the effects of the
channels explicitly included in our calculations\cite{and}.

The number of two- and three-phonon states constructed out of the
one-phonon basis given in table~\ref{basis-ca} is more than one
thousand. Considering that the amplitudes are complex quantities and
that we have an equation for each angular momentum and its projection,
the number of time dependent coupled equations to solve amounts to
about ten thousand. We have then to reduce their number in order to
render the calculation feasible. We took into account only the natural
parity states and furthermore we have considered for the calculations
only states with an excitation energy below 60 MeV. This cut off in
the excitation energies guarantees that we take into account almost
all the two-phonon states and a great number of the three-phonon
ones. Furthermore, we took into account, for each state, only the
components whose value is larger than 0.03. This choice guaranties
still a very good normalization and reduces appreciably the computation
time.


\begin{figure}[b!]
\begin{center}
\includegraphics[angle=0, width=1.0\columnwidth]{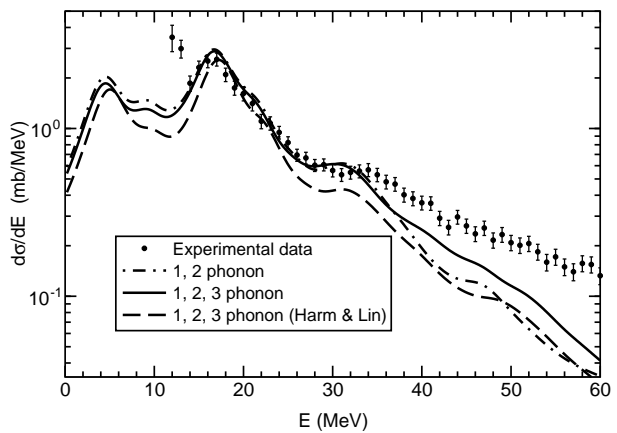}
\end{center}
\caption{
\label{fig:Ca-3pho-vs-2pho}
Comparison of the cross section for $^{40}Ca$+$^{40}Ca$ at 50 MeV/A
computed including only 1 and 2 phonons (dot-dashed line) with the
complete calculation, namely with 1,2 and 3 phonons (solid line). The
dashed line correspond to the complete calculation in the harmonic and
linear case. For the experimental data see ref.\cite{and}.}
\end{figure}
The contribution of the three-phonon states to the inelastic cross
section for $^{40}Ca$+$^{40}Ca$ at 50 MeV/A can be appreciated in
figure \ref{fig:Ca-3pho-vs-2pho} where the results of the calculations
in the larger space (solid line) are compared with the two-phonon ones
(dot-dashed line)\cite{and}.  Actually, we calculated the inelastic
cross section for each individual state (\ref{mix-3}) by solving the
coupled channel equations (\ref{chan}). The curves presented here are
always the result of a smoothing procedure with a Lorentzian with a
width of $\Gamma$=5 MeV (7 Mev for excitation energies greater than 30
MeV) of the theoretical cross sections to the discrete levels.  It
appears that both the one-phonon and the two-phonon strengths are a
little bit influenced by the inclusion of the three phonon
states. Although the three-phonon configurations appreciably affect
both the energies and wave functions of the two-phonon
states\cite{fal}, their role in the calculation of the cross section
seems to be small. On the contrary, in the three-phonon region the
increase of the cross section is appreciable and one can see that the
inclusion of the three phonon components improves the agreement with
the experimental data although there is still some cross section
missing. One might argue that a further enlargement of the
diagonalization space by including up to four-phonon states could
reduce this discrepancy. As we will see in the next section this is
not the case. Therefore some processes which are not taken into
account in our approach might be present in this energy region.
Indeed, the experimental spectrum compared with the calculations is an
excitation energy spectrum of $^{40}$Ca in coincidence with only one
detected proton at backward angles.  This coincidence measurement
makes sure that below 35 MeV no reaction mechanism participates to the
inelastic channel. However, above this excitation energy, a second
particle can be emitted in the forward direction through a different
reaction mechanism (such as the towing mode), leading to a higher
cross section than for a simple excitation. This can be evidenced by
means of velocity plots [thesis Mumu] that clearly show an asymmetry.
Eventhough one proton is emitted backward another one can be emitted
forward contributing to an increased cross section around 40 MeV.

In figure \ref{fig:Ca-3pho-vs-2pho} it is also plotted a curve
(dashed) which corresponds to the complete calculation in the harmonic
and linear case. Here, once again, we would like to underline the
importance of the anharmonic and non linear contributions. Indeed,
from the figure one infers that their presence increases the cross
section along the whole energy range of the calculation, expecially in
the double and triple GR energy regions. The increase of the cross
section in the triple GR energy region could be thought to be due only
to the presence of many more states in that region when three-phonon
states are included. The comparison with the results in the harmonic
and linear limit, where we have the same number of states, puts in
evidence the real origin of the strong increase. Namely, as already
stressed in ref.\cite{lan1,and,lan2}, the non linearities open new
routes to the excitation of multiphonon states while the
anharmonicities allow to populate them through their one- and/or
two-phonon components.


\begin{figure}[b!]
\begin{center}
\includegraphics[angle=0, width=1.0\columnwidth]{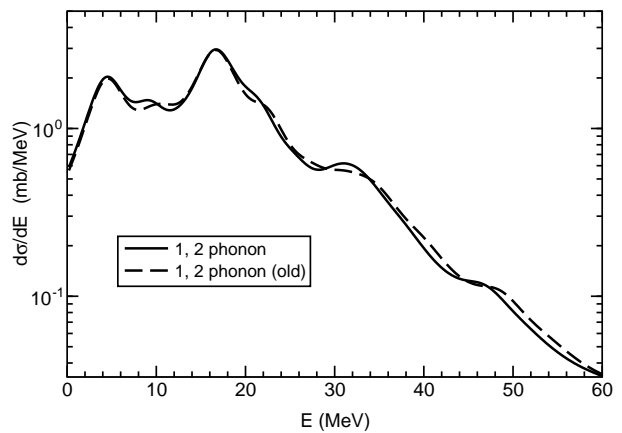}
\end{center}
\caption{
\label{fig:Ca-sigma-old-new}
Comparison of the inelastic cross-section for $^{40}Ca$+$^{40}Ca$ at
50 MeV/A between a calculation including only one- and two-phonon
states but with the energies and wave functions coming from the
diagonalization in the space of up to three-phonon states (solid line)
and the old two-phonon calculations (dashed line). }
\end{figure}
In figure \ref{fig:Ca-sigma-old-new} we compare the old two-phonon
calculation of ref.\cite{and} with the one done including only one-
and two-phonon channels but with the energies and wave functions
coming from the diagonalization of the Hamiltonian (\ref{h4}) in the
space of up to three-phonon states. We note that the different
calculations produce very similar results. The single resonance peak
is exactly the same in energy, intensity and shape while a little bit
of difference can be noted in the two-phonon region as well as in the
high energy region. The main effect, in these regions, is a small
shift towards lower energies of the whole curve which is due to the
relatively strong anharmonicities we have found when the three-phonon
states are included in the diagonalization. In any case, the two
curves are almost indistinguishable also in comparison with the
experimental data.

This study shows that the calculation of the cross section can be
considered to be converged at the n-phonon level when the (n+1)-phonon
states are taken into account in the structure calculation but are
neglected in the dynamical excitation.  In the next section we will
apply this recipe to the three-phonon excitation i.e. we include the
four-phonon configurations in the structure but not in the coupled
channel calculations.


\begin{figure}[b!]
\begin{center}
\includegraphics[angle=0, width=1.\columnwidth]{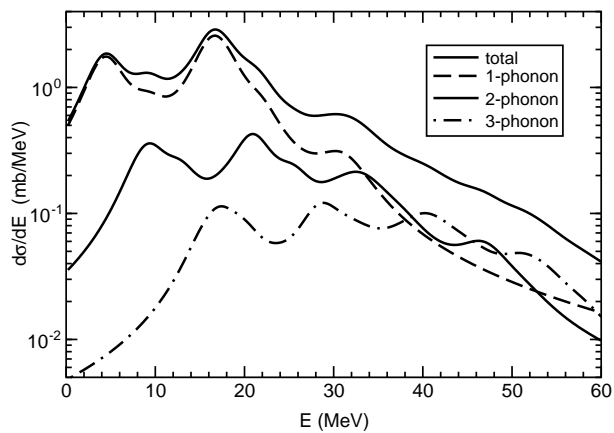}
\end{center}
\caption{
\label{fig:Ca-sigma-123pho}
Decomposition of the total inelastic cross-section (solid line, upper
one) for $^{40}$Ca + $^{40}$Ca at 50 MeV/A into the one- (dashed
line), two- (solid, lower one) and three-phonon components (dot-dashed
line).  }
\end{figure}

\begin{figure}[b!]
\begin{center}
\includegraphics[angle=0, width=1.2\columnwidth]{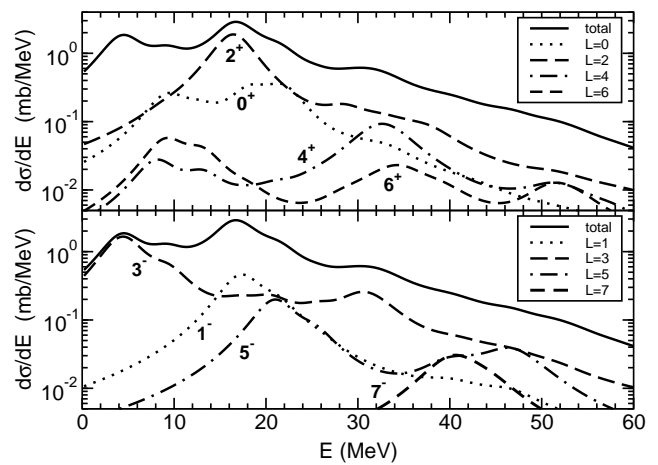}
\end{center}
\caption{
\label{fig:Ca-3pho-vs-l}
Decomposition of the total inelastic cross-section (solid line, in
both parts) for $^{40}Ca$+$^{40}Ca$ at 50 MeV/A into different angular
momenta contributions.  The figure is divide in two for the reader
convenience. In the upper part we show the even natural parity ones:
L=0 (dotted), L=2 (long-dashed), L=4 (dot-dashed) and L=6
(short-dashed).  In the lower part we plot the odd natural parity
angular momenta contribution: L=1 (dotted), L=3 (long-dashed), L=5
(dot-dashed) and L=7 (short-dashed).}
\end{figure}

To get a deeper insight in the role of the three phonon states, in
figure \ref{fig:Ca-sigma-123pho} we have decomposed the inelastic
cross-section into the one-, two- and three- phonon components.
The one-phonon distribution is dominated by the low-lying states and
the giant quadrupole resonance. The small peak at around 30 MeV
corresponds to the excitation of the HEOR state. All these states
are excited dominantly at low impact parameter through the nuclear
interaction. The incident energy and the projectile charge are not
large enough to induce a strong Coulomb excitation, so the GDR cross
section is four times smaller than the GQR one.  The two-phonon
contribution appears to be rather strong, about one order of magnitude
lower than the single phonon component. Its structure is more complex,
the various bumps being related to the double low-lying state
excitation, the excitation of a GQR on top of the low-lying mode, the
double GQR and the L=5 component of the $|\rm GQR\times \rm HEOR>$
state in the high energy tail. The latter contribution is also visible
in figure \ref{fig:Ca-3pho-vs-l}. The double GQR clearly dominates the
inelastic spectrum around 35 MeV excitation energy.

The three phonon component appears to be important. It is smaller than
the two-phonon strength by a factor of about 3. In the high energy
part it becomes the dominant contribution with a structure due to the
excitation of a low-lying mode on top of the double GQR state and
above 50 MeV to the triple GQR phonon. By inspection one can infer the
difference from the old calculation where only one- and two-phonon
states were taken into account. In table \ref{Sigma-Ca} we show the
summed cross sections in the 3 indicated regions, around the energies
corresponding to the excitation of one, two and three GQR phonons,
respectively. One can see that the cross sections corresponding to the
pure GQR states (within parentheses) decrease by a factor of about 10,
each time a new phonon is excited. For comparison, we show also, in
the second row, the results corresponding to the calculations done
with only one- and two-phonon states.

Because of the complex structure of both the two- and three- phonon
strength the total cross-section appears rather smooth above the
double GQR bump. However, the large cross-section makes possible
hunting for multiphonon states using selective signals such as
specific decays or multipolar decomposition.

\begin{table}[htdp]
\begin{center}
\caption{
  \label{Sigma-Ca} Integrated cross section (in mb) for $^{40}$Ca +
  $^{40}$Ca at 50 MeV/A in different energy bins corresponding to
  one-, two- and three-phonon regions, respectively.  In the first row
  there are shown the results for the three-phonon complete
  calculations. In the second row the results corresponding to the
  calculations done including only one and two-phonon states. In
  parenthesis the values corresponding to the single, double and
  triple GQR states.}
\smallskip
\begin{tabular}{|c|c|c|c|}
\hline
&(14-20 MeV)&(28-38 MeV)&(38-60 MeV)\\
\hline
3-pho\ \ \ \ & 20.83 (14.71)&  5.44 (1.34) & 2.02 (0.20)\\
\hline
2-pho\ \ \ \ & 22.80 (15.62)&  5.48 (1.97) & 0.66 (--)\\
\hline
\end{tabular}
\end{center}
\end{table}


The decomposition of the inelastic cross-section into various
multipoles is presented in figure \ref{fig:Ca-3pho-vs-l} where we show
the most significant contributions. In order to avoid an unreadable
figure, full of lines, we have separated it in two parts. In the upper
one, we show the even natural parity multipoles contributions while in
the lower part we plot the odd natural parity ones. The total cross
section is plotted in both graphs. The contribution corresponding to
the angular momenta L=0 and L=1 present appreciable peaks only in
correspondence of the single GMR and GDR, with a contribution of the
double 3$^-$ (L=0 component) for the monopole case. The single $3^-$
collective state dominates the low energy region. Comparing this
figure with the multiphonon decomposition one can deduce that the bump
in the $3^-$ strength at high energy, around 30 MeV, i.e. in the DGR
region, is due to the HEOR state.  The shoulder at higher energy is a
mixing of 2 and 3 phonon states. The quadrupole strength presents a
strong peak due to the GQR. The higher energy structures are mainly
due to the DGQR state except at very high energy, above 50 MeV, which
is dominated by the 3-GQR multiplet as can be inferred by the presence
of the same structure in the $2^+$, $4^+$ and $6^+$ components. The
strong peak in the L=5 contribution is due to the double phonon state
$|\rm GQR\times \rm 3^->$ while the one at higher energy correspond to
the excitation of the $|\rm GQR\times \rm HEOR>$ state. The 7$^-$
strength is due to the double GQR build on top of the low lying 3$^-$
state.

Since the presented strength is built from the monopole, dipole,
quadrupole and octupole collective states, the three-phonon component
goes up to L=9. However, it appears that the angular momenta above
L=7, which require three-phonon excitations, are a minor contribution
to the total strength. 


\subsection{ Role of 4 phonons }

\begin {table} 
\caption {Results of the diagonalization for
some two-phonon states of $^{40}$Ca. In the first column, the states
are labelled by their main component in the eigenvector and the
corresponding unperturbed energy, indicated in parentheses. In the
second column, the amplitude of the main component $c_0$. Then for
each total angular momentum, we show the results of the calculation in
the basis up to 2 phonon states, the results for the basis extended to
3 phonon states and the results for the basis up to 4 phonon
states. The last column contains the results in second order
perturbation theory. All energies are given in MeV.}
\smallskip
\label{ca40.res}
\begin{tabular}{lcccrcc}
\small Main & \small $c_0$ & \small $J^\pi$ & \small$\le2ph$ &
\small$\le3ph$ & \small$\le4ph$ & \small$2^{nd}$\\ 
\small component& $$ & $$ & $$ & $$& $$ & \small$order$ \\ 
\colrule
\small$3^-\!\!$$~\otimes$$\!\!~~3^- $
               &\small$-0.91$&\small$0^+$&\small$10.96$&\small$~9.27$&\small$~8.77$&\small$9.20 $\\
\small$(~9.88)$&\small$-0.96$&\small$2^+$&\small$10.63$&\small$~8.89$&\small$~8.43$&\small$~8.75$\\
               &\small$-0.96$&\small$4^+$&\small$~9.85$&\small$~8.10$&\small$~7.64$&\small$~7.96$\\
               &\small$-0.96$&\small$6^+$&\small$10.88$&\small$~9.12$&\small$~8.67$&\small$~8.99$\\ \hline

\small $D_1\!\!$$~\otimes$$\!\!~~D_1 $
               &\small$-0.92$&\small$0^+$&\small$35.27$&\small$33.71$&\small$33.35$&\small$33.59$\\
\small$(35.56)$&\small$-0.96$&\small$2^+$&\small$35.10$&\small$33.66$&\small$33.33$&\small$33.59$\\ \hline

\small $D_1\!\!$$~\otimes$$\!\!~~Q_1 $
               &\small$~0.95$&\small$1^-$&\small$34.83$&\small$33.35$&\small$33.05$&\small$33.24$\\
\small$(34.69)$&\small$~0.96$&\small$2^-$&\small$34.56$&\small$33.22$&\small$32.92$&\small$33.16$\\
               &\small$-0.96$&\small$3^-$&\small$34.67$&\small$33.13$&\small$32.82$&\small$33.02$\\ \hline

\small $Q_1\!\!$$~\otimes$$\!\!~~Q_1 $
               &\small$-0.87$&\small$0^+$&\small$33.88$&\small$32.47$&\small$32.03$&\small$32.27$\\
\small$(33.82)$&\small$~0.84$&\small$2^+$&\small$33.82$&\small$32.47$&\small$32.01$&\small$32.26$\\
               &\small$~0.90$&\small$4^+$&\small$34.02$&\small$32.61$&\small$32.18$&\small$32.44$\\ \hline

\small $M_2\!\!$$~\otimes$$\!\!~~D_1 $
               &\small$-0.89$&\small$1^-$&\small$40.26$&\small$38.14$&\small$37.72$&\small$37.65$\\
\small$(40.25)$&\small$     $&\small$   $&\small$     $&\small$     $&\small$     $&\small$     $\\ \hline

\small $M_2\!\!$$~\otimes$$\!\!~~Q_1 $
               &\small$-0.73$&\small$2^+$&\small$39.62$&\small$37.34$&\small$36.50$&\small$36.80$\\ 
\small$(39.38)$&\small$     $&\small$   $&\small$     $&\small$     $&\small$     $&\small$     $\\ \hline

\small $M_2\!\!$$~\otimes$$\!\!~~M_2 $
               &\small$~0.67$&\small$0^+$&\small$45.60$&\small$42.76$&\small$41.15$&\small$41.18$\\ 
\small$(44.94)$&\small$     $&\small$   $&\small$     $&\small$     $&\small$     $&\small$     $\\ \hline

\end{tabular}

\end{table}

The results from the previous calculation of anharmonicities evaluated
in a basis including up to three phonon states give us a clear
indication: it is not possible to compute the energies of three-phonon
states without including four-phonon states in the basis.  This
extension of the basis should be sufficient if we consider the weak
influence of the three-phonon states upon the one-phonon
ones. Moreover the results are well reproduced by second order
perturbation theory.  The introduction of four-phonon states in the
calculation could allow to test the convergence of the series, and to
study their effects upon one- and two-phonon states. Indeed, the
3-phonon states would undergo a strong energy shift towards low
energies, and their influence on 2-phonon states could be
substantially modified. We extend our basis to 4-phonon states to
compute anharmonicities and we will look at the effects on the
three-phonon states.


In order to do that, we follow the same approach described before: The
quartic Hamiltonian matrix (\ref{h4}) is diagonalized in the space of
up to four-phonon states, thus obtaining the new mixed eigenvectors:
\begin{eqnarray}\nonumber
|\Phi_\alpha> &=& \sum_{\nu_1} c_{\nu_1}^\alpha
|\nu_1> + \sum_{\nu_1 \nu_2}
c_{\nu_1 \nu_2}^\alpha |\nu_1\nu_2> \\ 
&+& \sum_{\nu_1\nu_2 \nu_3} c_{\nu_1
\nu_2 \nu_3}^\alpha |\nu_1\nu_2 \nu_3 > \\ \nonumber
&+&  \sum_{\nu_1\nu_2 \nu_3 \nu_4}
c_{\nu_1\nu_2 \nu_3 \nu_4}^\alpha |\nu_1\nu_2 \nu_3 \nu_4>
\end{eqnarray}

In Table \ref{ca40.res} we report the results of this calculation for
some relevant two-phonon states. Looking at the new energies, in the
sixth column, we remark that the additional shift imparted to the
2-phonon states by the inclusion of the 4-phonon states is smaller
than the previous case. Moreover the new shift is always towards lower
energies.

\begin {table*} [htb]
  \caption {Results of the diagonalization in the space including up
    to 4-phonon states.  The first column contains the name of the
    main component of the eigenvector. The harmonic energy of this
    state is written below the name. We show also the value of the c
    coefficient of the main component (second column) of the state as
    well as its parity and total angular momentum (third column). In
    the fourth and fifth columns we show the eigenenergies obtained
    when the diagonalization is done in a space up to three-phonon and
    up to four-phonon, respectively. These results have to be compared
    with the result of a second order perturbation theory calculation
    (sixth column). Finally, in the last two columns, there are shown
    other important components and the corresponding c coefficient.}
\smallskip
\label{ca40.res3}
\begin{tabular}{lccccrcc}
 Main &  $c_0$ &  $J^\pi$ & $\le3ph$ & $\le4ph$ & $2^{nd}$ &  important &  $c_i $ \\
 component& $$ & $$ & $$ &$$ &  order & components & $$ \\ 
\hline
$M_1 \otimes M_1 \otimes M_1$
               &-0.499 & $0^+$ & 54.48 & 53.12& 50.47
                           & $M_1\otimes3^-\otimes3^-$ &~-0.42\\
 54.74                &&&&&& $M_1 \otimes 3^- \otimes O_1$            &~0.41\\
                           &&&&&& $M_1 \otimes M_1 \otimes M_2$            &~0.26\\
                           &&&&&& $M_1 \otimes M_1 \otimes M_1 \otimes M_1$&~0.22\\
\hline
 $(D_1 \otimes D_2)_2 \otimes Q_1$
               &-0.37&$2^+$&56.17 &53.37&53.09
                           & $(D_1 \otimes D_2)_1 \otimes Q_1$&~0.32\\
 56.72                &&&&&&$(D_1 \otimes Q_2)_2 \otimes O_1$&~-0.19\\
                           &&&&&& $(D_1 \otimes D_2)_2 \otimes (3^- \otimes O_1)_2$&~-0.21\\
                           &&&&&& $(3^- \otimes 3^-)_2 \otimes (M_2 \otimes Q_1)_2$&~0.13\\
                           &&&&&& $(Q_1 \otimes Q_1)_0 \otimes (O_1 \otimes O_1)_2$&~0.36\\
                           &&&&&& $(Q_1 \otimes Q_1)_2 \otimes (O_1 \otimes O_1)_0$&~0.34\\
                           &&&&&& $(Q_1 \otimes Q_1)_4 \otimes (O_1 \otimes O_1)_2$&~0.36\\
                           &&&&&& $(Q_1 \otimes Q_1)_2 \otimes (O_1 \otimes O_1)_4$&~0.31\\
                           &&&&&& $(Q_1 \otimes Q_1)_2 \otimes (O_1 \otimes O_1)_2$&~-0.20\\
\hline
 $Q_1 \otimes Q_1 \otimes Q_1$
               &-0.91 &$0^+$& 50.74 & 47.8& 47.3 
                           & $(Q_1 \otimes Q_1)_2 \otimes (3^- \otimes 3^-)_2$&~0.24\\
 50.73                &&&&&& $Q_1 \otimes Q_1 \otimes Q_1 \otimes M_2$ &~-0.27\\
\hline
 $Q_1 \otimes Q_1 \otimes Q_1$ 
               &-0.91 & $2^+$& 50.96 & 48.0& 47.5 
                           &$(Q_1 \otimes Q_1)_0 \otimes (3^- \otimes 3^-)_2$&~-0.14\\
 50.73                &&&&&& $Q_1 \otimes Q_1 \otimes Q_1 \otimes M_2$& ~-0.27\\
                           &&&&&& $Q_1 \otimes Q_1 \otimes Q_1 \otimes M_1$&~-0.17\\
\hline
 $Q_1 \otimes Q_1 \otimes Q_1$
               &-0.65&$4^+$& 51.02 & 48.0& 47.56 
                           & $3^- \otimes 3^- \otimes 3^- \otimes O_2$ &~-0.18\\
 50.73                &&&&&& $Q_1 \otimes Q_1 \otimes Q_1 \otimes M_2$&~-0.19\\
                           &&&&&& $Q_1 \otimes Q_1 \otimes Q_1 \otimes M_1$ &~-0.13\\
                           &&&&&& $(Q_1 \otimes Q_1)_4 \otimes (3^- \otimes 3^-)_2$&-0.12\\
                           &&&&&& $ M_2 \otimes D_1 \otimes (Q_2 \otimes 3^-)_3$ &~0.18\\
                           &&&&&& $ M_1 \otimes D_1 \otimes (Q_2 \otimes 3^-)_3$ &~0.14\\
\hline
 $Q_1 \otimes Q_1 \otimes Q_1$
               &-0.92&$6^+$& 51.34 & 48.3& 47.85 
                           & $ Q_1 \otimes Q_1 \otimes Q_1 \otimes M_2$&~-0.27\\
 50.73                &&&&&& $(Q_1 \otimes Q_1)_4 \otimes (3^- \otimes 3^-)_2$&-0.20\\
                           &&&&&& $Q_1 \otimes Q_1 \otimes Q_1 \otimes M_1$&~-0.18\\
\hline
\end{tabular}
  
\end{table*}

The characteristics of a few 3-phonon states computed in the new basis
are presented in Table \ref{ca40.res3}. Anharmonicities of 3-phonon
states are still well reproduced by second order perturbation theory
and the correction to the harmonic energy is still negative. The
energy shift is due to the presence of the 4-phonon states which push
downwards the 3-phonon ones. This can be understood looking at the
second order correction to the energy in perturbative theory, where
the ratio between the matrix elements of the residual interaction
appearing at the numerator and the difference in energy at the
denominator determine the properties of the state. The largest matrix
elements are those coupling 3-phonon states to 4-phonon ones. This is
especially true in the cases involving triple and quadruple GMR states
and arises from symmetry properties of the phonons, obeying the Bose
statistics. In general, similarly to the findings of ref.\cite{fal},
the matrix elements connecting a n-phonon state to that formed by
adding to it a GMR are large. The sign of the correction to the
energy comes from the denominator, i.e. the difference between the
energy of the considered state and the 4-phonon states, the latters
being nearly all located at higher energies.  

The presence of the four-phonon states in the diagonalization basis
generates eigenfunctions that are more mixed than the ones of the
previuos calculations. So we get states having a main component of the
order of 0.5 and several others almost as large as it. Some examples
are given in Table \ref{ca40.res3}. The extreme case is the 2+ state
at 53.37 MeV excitation energy, whose main component is $(D_1 \otimes
D_2)_2 \otimes Q_1$ and appears with an amplitude of -0.37 in the
wavefunction. This state has several other components as large as
that, which are made of 3 and 4 phonons. 

The same is true for the L=4 state whose main component is $Q_1
\otimes Q_1 \otimes Q_1$ with an amplitude of -0.65. The same state is
much less mixed when the diagonalization is done in the space up to
three-phonon states. In the latter case its main coefficient is 0.986
with only one big component corresponding to the state $(D_1 \otimes
D_2)_2 \otimes Q_1$ whose amplitude is $c=0.15$. Another interesting
result, already found in \cite{fal} and related to the strong coupling
regime, is the existence of some states having as second large
component a configuration which is not directly coupled to the main
one by the residual interaction. This is a second order effect due to
the fact that both these configurations have large matrix element with
another one.

In order to check the stability of the results on the inelastic
scattering cross sections to two- and three-phonon states, we have
repeated the calculations by using the energies and wavefunctions
obtained by diagonalizing the hamiltonian in the large space but not
including the four-phonon channels. We dot not show the results of
this calculations because they are almost indistinguishable from the
previous ones. Once again, the inclusion of the n+1 phonons is very
important in the structure of the n-phonons but it seems it does not
affect strongly the dynamics. Therefore we can conclude that
convergence has been reached (at least numerically) and the
discrepancy between theory and experiment in the three-phonon region
is due to the presence in the experimental data of some processes
which are not taken into account in our approach.


\section{Excitation of $^{208}Pb$ }

We have also performed calculations for multiple excitation of
$^{208}Pb$. The inelastic cross section for the system $^{208}Pb$ +
$^{208}Pb$ has been computed for an incident energy of 641 MeV/A. At
this energy the nuclear contribution is believed to be small, so only
the relativistic Coulomb excitation has been taken into account in the
same way as it is described in ref.\cite{lan1}. The collective RPA
basis states considered in the present calculation are listed in table
\ref{basis-pb}. As in the previous case we construct all the possible
two- and three-phonon states and we diagonalize the Hamiltonian in
this space. For this case we will not take into account the effects of
the four-phonon states.

\begin{table}[htdp]
\begin{center}
\caption{Same as table \ref{basis-ca} for $^{208}$Pb.}
\label{basis-pb}
\smallskip
\begin{tabular}{|l|crrrr|}
\hline
State&$J^\pi$&$E_{harm}$ & $EWSR$&$E_{2ph}$&$E_{3ph}$\\ 
& &$(MeV)$ &  $(\%)$&$(MeV)$&$(MeV) $\\ 
\hline
GMR$_1$ & $0^+$ &$ 13.61  $&$ 61$&$ 13.42$&$ 13.48 $\\       
GMR$_2$ & $0^+$ &$ 15.02  $&$ 28$&$ 14.78$&$ 14.76 $\\ \hline 
GDR$_1$ & $1^-$ &$ 12.43  $&$ 63$&$ 12.30$&$ 12.30 $\\    
GDR$_2$ & $1^-$ &$ 16.66  $&$ 17$&$ 16.61$&$ 16.60 $\\ \hline
$2^+$   & $2^+$ &$ ~5.54  $&$ 15$&$ ~5.18$&$ ~5.14 $\\     
ISGQR   & $2^+$ &$ 11.60  $&$ 76$&$ 11.59$&$ 11.55 $\\      
IVGQR   & $2^+$ &$ 21.81  $&$ 45$&$ 21.69$&$ 21.68 $\\ \hline
$3^-$   & $3^-$ &$ ~3.46  $&$ 21$&$ ~3.21$&$ ~3.19 $\\      
HEOR    & $3^-$ &$ 21.30  $&$ 37$&$ 21.19$&$ 21.20 $\\
\hline
\end{tabular}
\end{center}
\end{table}
%
%
\begin{figure}[]
\begin{center}
\includegraphics[angle=0, width=1.\columnwidth]{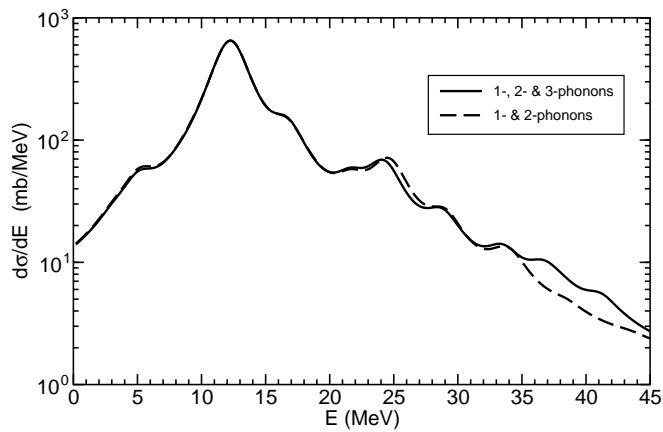}
\end{center}
\caption{\label{fig:Pb-sigma-old-new}
Comparison of the inelastic cross-section for $^{208}Pb$+$^{208}Pb$ at
641 MeV/A computed in ref. \cite{lan1} including only one- and
two-phonon states (dashed line) and the complete calculation going up
to three-phonons (solid).  }
\end{figure}
In figure \ref{fig:Pb-sigma-old-new} we compare the complete
calculation going up to three- phonons (solid line) with the
previously published\cite{lan1} inelastic cross-section for
$^{208}Pb$+$^{208}Pb$ at 641 MeV/A where only one- and two- phonon
states were included. One can see that below 35 MeV the results are
not affected by the inclusion of three-phonon states. Only a small
reduction of the peak around 25 MeV, corresponding to the double
GDR, is visible. This reduction can be related to the feeding of the
three phonon region, possible in the new calculations.
%
%
\begin{figure}[]
\begin{center}
\includegraphics[angle=0, width=1.\columnwidth]{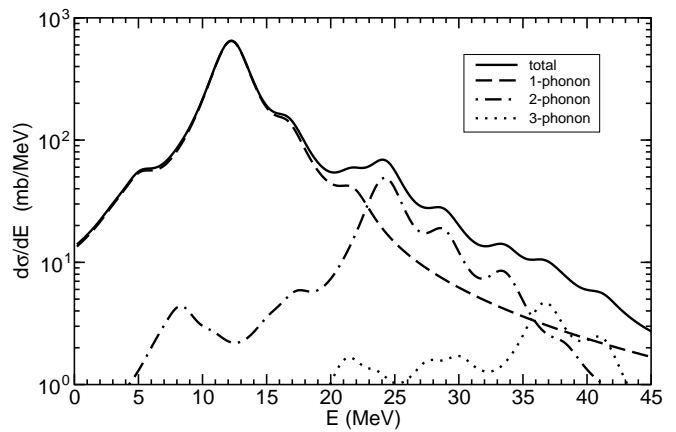}
\end{center}
\caption{\label{fig:Pb-sigma-123pho}
Decomposition of the inelastic cross-section for $^{208}Pb$+$^{208}Pb$
at 641 MeV/A (solid line) into the one- (dashed), two- (dot-dashed) and
three-phonon components (dotted).  }
\end{figure}

In the high energy region the contribution of the three phonon states
appears to be important. This is confirmed by the decomposition of the
inelastic cross-section into the one- , two- and three-phonon
components as shown in figure \ref{fig:Pb-sigma-123pho}. At this
relativistic energies and with such heavy charged ions the one-phonon
cross-section is clearly dominated by the GDR. In the figure, the
small shoulder at 17 MeV is due to the high lying component of the GDR
carrying a small fraction of the dipole strength. The one-phonon
component around 22 MeV is the isovector quadrupole vibration. The
double phonon component is clearly dominated by the double-GDR
excitation. In fact from 23 to 34 MeV the inelastic cross-section
appears to be mainly due to two-phonon states. Indeed, the first peak
in this energy region corresponds to the L=2 component of $|GDR_1
\times GDR_1>$, the second peak is the L=2 component of $|GDR_1 \times
GDR_2>$ and the third one is the L=3 component of $|GDR_1 \times
IVGQR>$. The double $GDR_2$ state has a small cross section and it
cannot be appreciated in the figure. Above 35 MeV the three-phonon
modes provide the most important contribution to the
spectrum. Indeed, the main peak is due to the triple $DGR_1$ while the
second one corresponds to the $|GDR_1\times GDR_1 \times GDR_2>$ state.

\begin{table}[htdp]
\begin{center}
\caption{
\label{Sigma-Pb}
Integrated cross section (in mb) for $^{208}Pb$+$^{208}Pb$ at 641
MeV/A in different energy bins corresponding respectively to the GDR,
the two-phonon and the three-phonon regions. In the second row the
calculations done with only one- and two-phonon states. In parenthesis
the values corresponding to the single, double and triple GDR states.}
\smallskip
\begin{tabular}{|c|c|c|c|}
\hline
&(8-19 MeV)&(22-35 MeV)&(35-45 MeV)\\
\hline
3-pho \ \ \ \ &3451.2 (3078.4)&325.6 (227.1)&39.2 (18.7)\\
\hline
2-pho \ \ \ \ &3510.3 (3103.3)&348.6 (245.0)&4.0 (--)\\
\hline
\end{tabular}
\end{center}
\end{table}%

From table \ref{Sigma-Pb} one can see that the integrated
cross-section for the excitation of the GDR is large at such a
relativistic energy, reaching 3.5 barns. Then a factor 10 has to be
paid each time a new phonon is excited still leaving some sizeable
cross-section for two and three phonon excitations. In the second row
we show the results for the calculations done with only one- and
two-phonon states. In the three-phonon region we gain a factor 10 when
we introduce in the calculation the three-phonon states.

\begin{figure}[]
\begin{center}
\includegraphics[angle=0, width=1.\columnwidth]{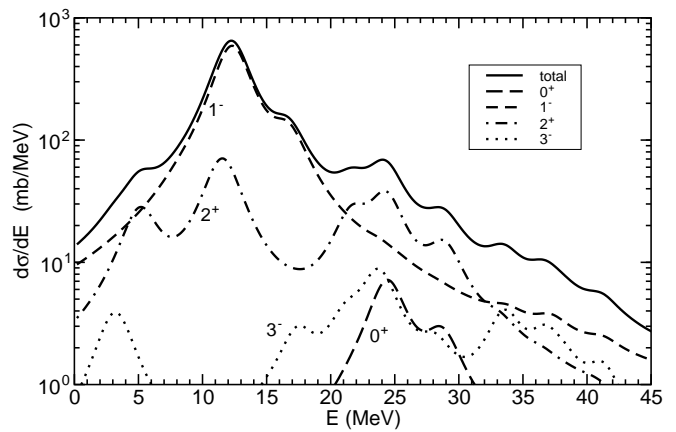}
\end{center}
\caption{\label{fig:Pb-3pho-vs-l}
Decomposition of the inelastic cross-section for $^{208}Pb$+$^{208}Pb$
at 641 MeV/A (solid line) into different angular momenta, L=0
(long-dashed), L=1 (short-dashed), L=2 (dot-dashed) and L=3
(dotted). }
\end{figure}

To get a deeper insight in the excitation process, it is interesting
to decompose the computed inelastic spectrum in various
multipolarities. Let us first start with the dipole strength which
strongly dominates at relativistic energies (see figure
\ref{fig:Pb-3pho-vs-l}). The GDR, which is splitted into a main
component at 12.5 MeV and a smaller peak around 17 MeV, is of course
the main contributor but one can observe above 35 MeV a small
contribution of the three-GDR state coupled to  spin and
parity $1^-$. The quadrupole strength is more complex. Starting at low
energy one observes the low lying collective $2^+$ state followed by
the isoscalar GQR just above 10 MeV. Except for the small shoulder at
22 MeV coming from the isovector GQR, the strong bump at 25 MeV can be
essentially attributed to the double GDR coupled to $2^+$. This peak
can be directly compared with the monopole strength which corresponds
entirely to the double GDR coupled to $0^+$. Coming back to the
quadrupole response one notices that the two phonon contribution
appears as strong as the one phonon excitation. A multipole analysis
can thus be an interesting way to experimentally isolate the
multiphonon contribution. Finally, the octupole response presents both
a $3^-$ and HEOR components around 3 and 24 MeV, followed by the triple
GDR state coupled to $3^-$ around 35 MeV. This three phonon component
corresponds to the structures observed in the $1^-$ response which is
nothing but the low spin member of the three-phonon multiplet.
%

\section{Conclusion }

In this paper we present, for the first time, microscopic calculations
of inelastic cross sections for the triple excitation of giant
resonances induced by heavy ion probes.

We use a microscopic approach based on RPA: the mixing of three-phonon
states among themselves and with two- and one-phonon states is
considered within a boson expansion approach with Pauli
corrections. This is equivalent to introduce anharmonicities
corrections to the standard harmonic approximations. At the same time
we have also introduced non-linearities in the external field.

The calculations were done by solving semiclassical coupled channel
equations, the channels being superpositions of multiphonon states. In
previous calculations we have considered only one- and two-phonon
states obtaining a good agreement with the experimental cross section.

In this paper we extend these microscopic calculations by including
the three-phonon states.  By diagonalizing a quartic microscopic
Hamiltonian in the space of up to three-phonon states one realizes
that a correct description of two-phonon states requires the inclusion
of one and three-phonon components. The anharmonicity in most of the
cases is of the order of 1 MeV. Calculations of the inelastic cross
section for the excitation of one-, two- and three-phonons states have
been performed in the framework of this model. The cross section in
the DGR energy region is only slightly modified. Thus the previously
published results are confirmed. On the contrary, as one could expect,
the contribution in the TGR energy region is quite large giving a
better agreement with the experimental data.  We have also performed
calculations in the space of up to four-phonon states. Although the
inclusion of the four-phonon states is very important in the wave
functions and energies of the three-phonon states, giving rise to a
much stronger anharmonicity, their influence on the dynamics is very
small.

The decomposition of the inelastic cross section into one-,
two- and three-phonon components shows the importance of the three
contributions in different region of the excitation energy. In the
case of $^{208}$Pb + $^{208}$Pb at E/A=641 the separation in energy is
very clear and one can distinguish the three region of interest, in
the $^{40}$Ca + $^{40}$Ca at 50 MeV/A case the overlap is stronger. In
both cases we get an increase in the triple phonon energy region
showing once again the importance of the anharmonicity in the internal
hamiltonian and the non-linearity in the external field.

\end{document}